\documentclass[aps, prb, onecolumn, groupedaddress, preprint]{revtex4-1}
\usepackage{graphicx}
\usepackage{epstopdf}
\usepackage{amsmath,amssymb,amsfonts}
\usepackage{color}
\newcommand{\fL}{\mathcal{L}}

\newcommand{\bC}{\mathbb{C}}

\newcommand{\be}{\begin{equation}}
\newcommand{\ee}{\end{equation}}
\newcommand{\tU}{\text{U}}

\newcommand{\tr}{\text{tr}}

\begin{document}

\title{Mass generation by fractional instantons in SU($n$) chains}
\author{Kyle Wamer and Ian Affleck}
\affiliation{ Department of Physics and Astronomy and Stewart Blusson Quantum Matter Institute, University of British Columbia, 
Vancouver, B.C., Canada, V6T1Z1}
\date{\today}      
 
\begin{abstract}

Recently, Haldane's conjecture about SU(2) chains has been generalized to SU($n$) chains in the symmetric representations. For a rank-$p$ representation, a gapless phase is predicted when $p$ and $n$ are coprime; otherwise, a finite energy gap is present above the ground state. In this work, we offer an intuitive explanation of this behavior based on fractional topological excitations, which are able to generate a mass gap except when $p$ and $n$ have no common divisor. This is a generalization of an older work in SU(2), which explains the generation of the Haldane gap in terms of merons in the O(3) nonlinear sigma model.

\end{abstract}

\maketitle

\section{Introduction}

Haldane's conjecture is the prediction that antiferromagnetic spin chains with integer spin have a gap above the ground state, while those with half-odd integer spin are gapless.\cite{HaldanePRL1983, HaldanePLA1983} The distinction between these two cases can be seen by taking a large spin limit, in which case the quantum fluctuations of the antiferromagnet are governed by the O(3) nonlinear sigma model, with topological angle $\theta= 2\pi s$:
\be \label{eq:1}
	\fL = \frac{1}{2g}(\partial_\mu \vec{n})^2 +\fL_{\text{top}}, \hspace{10mm}  \int d^2x \fL_{\text{top}} \in \theta \mathbb{Z}.
\ee
This sigma model, in the absence of a topological term, is known to have a finite energy gap above the ground state.\cite{ZAMOLODCHIKOV1979253} In [\onlinecite{PhysRevLett.56.408}], it was argued that this energy gap can be thought of as being generated by half-quantized topological excitations called merons. Merons behave like planar vortices far from their centers, but lift out of the plane near their cores, and form finite action configurations when a collection of them has a net vorticity of zero. Remarkably, when the O(3) sigma model has a topological angle of $\pi$ (corresponding to a half-odd-inetger spin chain), the merons that point up at their origin destructively interfere with the merons that point down at their origin (also known as antimerons), and the mass generating mechanism of the model is no longer effective. While this isn't the full story, since larger-action configurations such as double-merons do not destructively interfere at $\theta=\pi$, it offers an intuitive picture in favor of Haldane's conjecture. 

Recently, this paradigm of mapping spin chains to relativistic quantum field theories has been generalized to SU($n$) chains in the rank-$p$ symmetric representations.\cite{BYKOV2012100, bykov2013geometry, LajkoNuclPhys2017,2019arXiv191008196W} In this case, the O(3) model is replace with a $\mbox{SU}(n)/[\tU(1)]^{n-1}$ flag manifold sigma model, which is described by $n$ orthonormalized fields $\vec{\phi}^\alpha \in \bC^n$, and has $n-1$ independent topological angles, $\theta_\alpha = \frac{2\pi p}{n} \alpha$.  According to [\onlinecite{2019arXiv191008196W}], these sigma models have a gapless ground state for all $p$, except in the special case when $p$ and $n$ have a common divisor greater than 1, in which case a gapped phase is possible. This is in accordance with the Lieb-Schultz-Mattis-Affleck (LSMA) theorem, which predicts either a gapless phase or spontaneously broken translation symmetry when $p$ is not a multiple of $n$.\cite{LSM1961, AffleckLieb1986} A subtle feature of this result, which can only occur for $n\geq 4$, is that when $p$ and $n$ have a nontrivial divisor, and $p$ is a not a multiple of $n$, translation symmetry is always spontaneously broken. This follows from an 't Hooft anomaly that forbids the flow of the above sigma model to SU($n)_1$, the only stable SU($n$) fixed point.\cite{Yao:2018kel, ohmori2019sigma, Tanizaki:2018xto}

In this article, we offer an intuitive explanation of this gap dependence on $p$ and $n$, based on a mass generation mechanism due to fractionalized topological excitations. It is the SU($n$) generalization of the meron story in the O(3) sigma model. In Section \ref{sec:rev}, we review the derivation of the flag manifold sigma model from the SU($n$) chain, following [\onlinecite{2019arXiv191008196W}], and discuss the various symmetries of the model. In Section \ref{sec:symbreak}, we add anisotropic potentials to the SU($n$) chain that break the continuous global symmetry down to U(1), and classify the topological excitations that have finite action. Next, in Section \ref{sec:massgen}, we show how these excitations, so-called fractional instantons which have topological charge $\frac{1}{n}$, are responsible for generating a mass gap in the sigma model, in a similar fashion as the vortices in the Kosterlitz-Thouless transition. \cite{Kosterlitz_1973} However, this mechanism is highly dependent on the topological angles of the model, which are determined by the representation $p$ of SU($n$). This dependence is discussed in Section \ref{sec:top}, where we ultimately conclude that mass generation by fractional instantons occurs except when $p$ and $n$ are coprime, in which case a gapless phase is possible. 
Our concluding remarks can be found in Section \ref{sec:con}.

\section{SU($n$) Chains and Flag Manifold Sigma Models: A Review} \label{sec:rev}

The SU($n$) generalization of the antiferromagnetic spin chain can be written in terms of $n\times n$ traceless matrices, $S$, whose components contain the $n^2-1$ generators of SU($n$):
\be \label{eq:ham}
	H = J\sum_j \tr[ S(j)S(j+1)].
\ee
These matrices satisfy the following commutation relations,
\be
	[S_{\alpha\beta}, S_{\mu \nu}] = \delta_{\alpha\nu} S_{\mu \beta} - \delta_{\mu\beta} S_{\alpha\nu},
\ee
and the eigenvalues of $S$ are constrained by the the SU($n$) representation that occurs at each site of the spin chain. Of primary interest are the chains that at each site have the same rank-$p$ symmetric representation, since these most naturally lead to a generalization of Haldane's conjecture (here $p$ is a positive integer). By adding longer range interactions{\begin{footnote}{One must add up to $(n-1)$th neighbour interactions in order to stabilize the classical ground state, and allow for a field theory analysis. It is believed that these longer range terms in the Hamiltonian will be generated dynamically from the original Hamiltonian (\ref{eq:ham}). See  [\onlinecite{2019arXiv191008196W}] for details.}\end{footnote}} and taking the large-$p$ limit, the matrices are forced to take the form
\be \label{eq:ss}
	S_{\alpha \beta}(j_\gamma) = p [\phi_\gamma^\alpha]^* \phi_\gamma^\beta.
\ee
Here we've defined $j_\gamma = nj + \gamma-1$, which captures the $n$-site order of the classical ground state. The vectors $\vec{\phi}^\alpha$ form an orthonormal basis of $\mathbb{C}^n$. By considering low energy fluctuations about this classical ground state, we obtain a quantum field theory in terms of these $\vec{\phi}^\alpha$. At low enough energies, this quantum field theory flows to a Lorentz invariant flag manifold sigma model, of the form $\fL = \sum_{\alpha<\beta} \fL_{\alpha\beta} + \fL_{\text{top}}$, where 
\be \label{eq:2}
	\fL_{\alpha\beta}= \frac{ \big|\vec{\phi}^\alpha \cdot \partial_\mu \vec{\phi}^{*\beta}\big|^2}{g_{\alpha,\beta}}
	+ \epsilon_{\mu\nu}\lambda_{\alpha,\beta} (\vec{\phi}^\alpha\cdot\partial_\nu \vec{\phi}^{*\beta} )(\partial_\mu \vec{\phi}^\beta \cdot\vec{\phi}^{*\alpha})
\ee
and $\fL_{\text{top}} = \sum_\alpha \theta_\alpha q_\alpha$, with $\theta_\alpha = \frac{2\pi p\alpha }{n}
$ and 
\be
	 q_\alpha = \frac{1}{2\pi i} \epsilon_{\mu\nu}\partial_\mu \vec{\phi}^\alpha \cdot\partial_\nu \vec{\phi}^{*\alpha},
	 \hspace{10mm}
	 \int d^2x q_\alpha \in \mathbb{Z}.
\ee
The coupling constants $g_{\alpha,\beta}$ and $\lambda_{\alpha,\beta}$ are functions of the antiferromagnetic interaction strength at the bare level. This model possesses a $[\tU(1)]^{n-1}$ gauge symmetry, corresponding to a local phase rotation for each of the $\vec{\phi}^\alpha$ (the $n$th field, $\vec{\phi}^n$, can be written in terms of the first $n-1$ fields, and does not admit its own, independent gauge symmetry). It also has a global $\mathbb{Z}_n$ symmetry,
\be
	\vec{\phi}^\alpha \mapsto \vec{\phi}^{\alpha+1},
\ee
which is the manifestation of the translation invariance that is present at the spin chain level. 

\section{Reducing the Global Continuous Symmetry to U(1) } \label{sec:symbreak}

Following [\onlinecite{PhysRevLett.56.408}], our strategy is to break the symmetry of the flag manifold down to U(1), where a phase transition is well understood in terms of vortex proliferation. We first add to (\ref{eq:2}) an anisotropic potential $V_1$ that breaks the SU($n$) symmetry down to $[\tU(1)]^{n-1}$, while preserving the discrete $\mathbb{Z}_n$ symmetry and the $[\tU(1)]^{n-1}$ gauge structure:
\be \label{eq:v}
	V_1 = m\sum_{\alpha=1}^n \sum_{\beta <\gamma} \big( |\phi^\alpha_\beta|^2 - |\phi^\alpha_\gamma|^2\big)^2.
\ee
This is the SU($n$) generalization of adding the term $\sum_j {S}_z(j){S}_z(j)$ to the SU(2) Hamiltonian. Indeed, using (\ref{eq:ss}), it is straightforward to rewrite $V_1$ in terms of the SU($n$) generators on the chain (see Appendix \ref{app:pot}). In the limit $m\to \infty$, the potential $V_1$ restricts each of the $\vec{\phi}^\alpha$ to lie on an $n-1$ torus, parameterized by $n-1$ angles, $\varsigma_\beta = \varsigma_\beta(\alpha)$:
\be \label{eq:4}
	\vec{\phi}^\alpha = \frac{1}{\sqrt{n}} \begin{pmatrix}  e^{i\varsigma_1} & e^{i\varsigma_2} & \cdots & e^{i\varsigma_{n-1}} & e^{i\varsigma_n}  \\
	\end{pmatrix}^\text{T}.
\ee

Throughout, we fix a gauge by choosing $\phi^\alpha_1 $ to be real and positive, for all $\alpha$. Since the number of free parameters equals the number of orthogonality constraints (both equal $n(n-1)/2$), one might conclude that the there is a unique configuration of the $\{\vec{\phi}^\alpha\}$ in this limit (up to permutations). However, the orthogonality constraints are invariant under the following transformations,
\be \label{eq:3}
	\phi^\alpha_\beta \to e^{i\theta_\beta}\phi^\alpha_\beta \hspace{10mm} \forall \alpha,\beta,
\ee
which are true symmetries for each of the $n-1$ un-gauge-fixed directions, $\beta = 2,3,\cdots, n$. Therefore, by introducing the potential (\ref{eq:v}), the continuous symmetry group of the sigma model has been broken down to $[\tU(1)]^{n-1}$. A typical configuration of the $\{\vec{\phi}^\alpha\}$ in this submanifold can be constructed from the $n$th root of unity, $\zeta := e^{i\frac{2\pi}{n}}$:
\be \label{eq:typical}
	\phi^\alpha_\beta = \frac{1}{\sqrt{n}}\zeta^{\alpha\beta} e^{i\sigma_\beta} \hspace{5mm} \sigma_\beta \in [0,2\pi].
\ee
Indeed, orthonormality of these states follows from the identity $
	\sum_{\alpha=1}^n \zeta^{\rho \alpha} =0$, which holds for all $\rho\not= 0 \mod n$. Since the Lagrangian (\ref{eq:2}) couples all of the $n-1$ copies of U(1) together, it will be useful to break the symmetry down further. To this end, we introduce a second potential $V_2$, given by\begin{footnote} {See Appendix \ref{app:pot} for the form that $V_2$ takes on the SU($n$) chain}\end{footnote}
\be \label{eq:4b}
	V_2=-m\sum_{\alpha=1}^n\sum_{\beta=2}^{n-1} \Big((\phi^\alpha_1[\phi^\alpha_\beta]^*)^n - ([\phi^\alpha_1]^*\phi^\alpha_\beta)^n\Big)^2.
\ee
The factors of $n$ in the exponent are necessary in order to preserve the $\mathbb{Z}_n$ symmetry, which corresponds to translational invariance in the underlying lattice model.{\begin{footnote} {Since $\vec{\phi}$ has dimension 0 in 1+1 dimensions, an operator with arbitrarily high powers of $\phi$ is still renormalizable.}\end{footnote}} Since $V_2$ is still minimized by the configurations (\ref{eq:typical}), it can be rewritten in terms of the U(1) fields, $\sigma_\beta$, yielding
\be
	V_2= 4m\sum_{\alpha=1}^n\sum_{\beta=2}^{n-1}\sin^2(n(\sigma_1-\sigma_\beta)).
\ee
It is clear that the effect of $V_2$ is to equate all but one of the U(1) fields. In our fixed gauge, this amounts to setting $\sigma_\beta =0$ for $\beta\not=n$, resulting in a theory that involves $\sigma_n$ only. This is equivalent to the O(2) model of a vector $\vec{n}\in\mathbb{R}^3$ restricted to the XY plane. By inserting this restricted form of $\vec{\phi}^\alpha$ into (\ref{eq:2}), it is easy to show that the $\lambda_{\alpha,\beta}$ terms vanish, and the resulting O(2) coupling constant is $g$, defined by
\be \label{eq:gg}
	g^{-1} = \sum_{\alpha,\beta} g_{\alpha,\beta}^{-1}.
\ee

\section{Mass Generation by Fractional Instantons} \label{sec:massgen}

In the absence of the $\theta_\alpha$, it is well known that a mass gap is generated in the O(2) model, due to vortex proliferation. Moreover, if we view the O(2) model as a special limit of the O(3) sigma model (\ref{eq:1}) with a large potential added to restrict the field $\vec{n}$ to the XY plane, this mass gap is still generated when the potential is weakened, and $\vec{n}$ can lift off the plane. Now, the gap is generated by nonplanar vortices, known as merons.\cite{PhysRevLett.56.408} This argument was used to identify merons as the mass-gap generating mechanism in the O(3) sigma model which corresponds to the purely isotropic case of $m=0$. Here, we would like to follow this same logic and argue that a mass gap is present in the SU($n)/[\mbox{U}(1)]^{n-1}$ flag manifold sigma model (without topological term), and that it is generated by topological excitations.

We begin by considering the case of $n=2$, where (\ref{eq:4b}) vanishes and the perturbation (\ref{eq:v}) is equivalent to adding a mass term $m(n_3)^2$ to the O(3) model (\ref{eq:1}), restricting $\vec{n}$ to lie in the XY plane in the large $m$ limit. This follows from the equivalence equation $n^i = \vec{\phi}^\dag \sigma^i \vec{\phi}$. While the lattice version of this theory experiences a phase transition due to vortex proliferation, in the continuum limit, the vortices must become non-planar at their core in order to avoid a UV singularity. It is these nonplanar vortices that are the merons, since they have topological charge $Q= \pm \frac{1}{2}$ (the sign depending on whether $n_3 = \pm 1$ at their core), as compared to $Q=\pm 1$ for the more familiar instantons and antiinstantons of the O(3) model.

To demonstrate the fractional nature of the meron charge, we return to the $\phi$ notation, since this will immediately generalize to the case of SU($n$). While there are two fields $\vec{\phi}^\alpha$ in SU(2), their mutual orthogonality allows us to restrict our attention to a single field, which we refer to as simply $\vec{\phi}$. The U(1) degree of freedom corresponds to the phase difference between the two components of $\vec{\phi}$. The meron configuration for $\vec{\phi}$ behaves like a traditional U(1) vortex far from its core, and has the form 
\be
	\vec{\phi} = \frac{1}{\sqrt{2}} \begin{pmatrix} 1 \\
	e^{i\omega} \\
	\end{pmatrix},
\ee
where $\omega$ is the polar angle in the XY plane. However, at the object's centre, this dependence on $\omega$ leads to a UV singularity, which can only be avoided by sending one of the two components of $\vec{\phi}$ to zero (which corresponds to $n_3 \to \pm 1$ in the O(3) model):
\be \label{eq:8}
	\vec{\phi}\to \begin{pmatrix} 1 \\
	0 \\
	\end{pmatrix} \hspace{5mm} \text{ or } \hspace{5mm}
	\vec{\phi} \to \begin{pmatrix} 0 \\
	e^{i\omega} \\
	\end{pmatrix}.
\ee
The phase remains in the second case due to our gauge choice; however, at the meron's core, it can be removed by a pure gauge transformation, and thus does not lead to a true UV singularity.

Since the topological charge, $Q$, of $\vec{\phi}$ is the integral of a total derivative, we may rewrite $Q$ as the difference of two contour integrals, one around the origin and one around infinity:
\be \label{eq:12}
	Q = \frac{1}{2\pi i}\left[ \oint_{0} dx^\mu  \vec{\phi} \cdot \partial_\mu \vec{\phi}^* - \oint_{\infty} dx^\mu \vec{\phi}\cdot\partial_\mu \vec{\phi}^*\right].
\ee
The contour is necessary to avoid any gauge singularity that may be present at the origin (see Appendix \ref{app:1}). For both possibilities in (\ref{eq:8}), the contour about infinity contributes a charge of $\frac{1}{2}$. Meanwhile, the contribution from the contour about the origin is trivial in the first possibility, and -1 in the second, leading to $Q = \pm \frac{1}{2}$, which proves the half-quantized charge of the meron.

Extending this calculation to SU($n$) is straightforward. Now, the topological charge is a vector $\vec{Q} = (Q_1,Q_2,\cdots, Q_n)$, corresponding to the $n$ fields $\vec{\phi}^\alpha$. A configuration far from its core has $\phi^\alpha_n = \frac{1}{\sqrt{n}}e^{i\omega}$ for all $\alpha$, and receives a contribution to the topological charge of $\frac{1}{n}$ from the contour at infinity. Meanwhile, in order to be UV finite at its core, the phase $e^{i\omega}$ in each of the $n$th components of the $\vec{\phi}_\alpha$ must either vanish or become a pure gauge transformation. This leads to an $n$-fold family of topological excitations, distinguished by which one of the $\vec{\phi}^\alpha$ tends to $(0,0,0,\cdots, e^{i\omega})^\text{T}$. For that particular $\vec{\phi}$, a contribution of -1 is added to its topological charge from the contour about the origin in (\ref{eq:12}), while the remaining $n-1$ fields receive no contribution, as they are topological trivially there. See Appendix \ref{app:2}) for an explicit solution in the case of $n=3$. Thus, we obtain $n$ species of topological configurations, with charges
\be \label{eq:qq}
	\vec{Q}^\beta = \frac{1}{n}\big( 1, 1, \cdots, \overbrace{-(n-1)}^{\text{position $\beta$}}, \cdots, 1 \big).
\ee
And so for all $n$, we find fractionalized topological excitations, or ``fractional instantons'', in our symmetry broken model.

While the number of such configurations has increased, the original argument from SU(2) for mass generation carries over to this more general case: for each species of topological excitation in this $n$-fold family, we have a species of particle in the Coulomb gas formalism.\cite{PhysRevLett.56.408} That is, each particle has a partition function that is represented (at large distances) by a sine-Gordon (sG) model,
\be \label{eq:SG}
	\fL_{SG} = \frac{1}{2} (\partial_\mu \sigma)^2 + \gamma \cos (\frac{2\pi}{g}  \sigma),
\ee
in the limit of large $\gamma$, which represents the fugacity, or density, of the fractional instanton gas. This expression is derived in detail in [\onlinecite{Affleck_1989}], and relies on the fact that all higher-loop corrections to the (fractional) instanton gas are IR finite in the sG model.\cite{Kosterlitz_1973} Here, $g$ is the O(2) coupling constant in (\ref{eq:gg}), and plays the role of temperature in the sG model. Since all of the $n$ species arise from the same action, each will have the same fugacity and critical $g$, so that the above model (\ref{eq:SG}) is merely copied $n$ times, and the SU(2) analysis from [\onlinecite{PhysRevLett.56.408}] can be applied directly: For large $m$, the fractional instantons are dilute and we are in a massless boson phase. As $m$ is lowered, the effective critical temperature is increased until the topological excitations condense and a mass gap is produced.\begin{footnote}{It is also worth noting that the exact critical exponents for the sG model are well known, and are reviewed in [\onlinecite{Affleck_1989}].}\end{footnote} Thus, we conclude that fractional instantons are responsible for generating a mass gap in the flag manifold sigma model (\ref{eq:2}), in the absence of topological angles.

\section{Destructive Interference in the Presence of Topological Angles} \label{sec:top}

We now restore the topological angles $\theta_\alpha$, and study how the mass generating mechanism changes. For large $m$, we are in the O(2) model and the $\theta$-terms do not play a role. However, as $m$ is lowered towards zero, the fugacity $\gamma$ in the sine-Gordon model is modified to
\be \label{eq:fug}
	\gamma\sum_{\beta=1}^n e^{i\vec{\theta}\cdot\vec{Q}^\beta},
\ee
where the sum is over the $n$ species of fractional instanton, and $\theta_\alpha = \alpha \theta$, with $\theta = \frac{2\pi p}{n}$. Using (\ref{eq:qq}), one easily finds that this sum equals $\gamma\sum_k \zeta^{pk}$. So long as $p$ is not a multiple of $n$, this sum vanishes. In other words, for such values of $p$, the fugacity of the sG model vanishes, suggesting that the Coulomb gas is always in its massless phase.

 At first glance, this appears to be inconsistent with the conjecture put forward in [\onlinecite{2019arXiv191008196W}], which also predicts a gap when $p$ is not a multiple of $n$, but has a nontrivial shared divisor with $n$. This discrepancy is resolved by considering higher order topological excitations. Indeed, while objects that have winding number greater than $\pm 1$ in $\omega$ have larger action, they too must be considered, and do not necessarily lead to a vanishing fugacity. A similar calculation shows that a configuration with winding number $j$ has charge $\vec{Q}^{\beta,j}$, with
\be
	Q^{\beta,j}_\alpha =\frac{j}{n} - j\delta_{\beta,\alpha},
\ee
and modifies the fugacity to $\gamma \sum_k\zeta^{pkj}$. If $j$ is a multiple of $n$, this will lead to a nonzero fugacity. This is also true in the case of SU(2); indeed, it was acknowledged in [\onlinecite{PhysRevLett.56.408}] that so-called ``double merons'' do not having cancelling contributions. However, the conclusion arrived at in that paper also applies here: since these higher vorticity objects have larger action, we can at least conclude that the critical value of $m$, above which occurs a gapless phase, is lower at $\theta=\frac{2\pi p }{n}$ than at $\theta=0$.

Now, if $\gcd(p,n)  \not= 1,n$, then not only do events with winding number $n$ contribute to the fugacity, but so do events with winding number $d := n/\gcd(p,n)$. Since these contributions will have a much smaller action than the $n$-winding events, this shows that whenever $p$ and $n$ have a nontrivial common divisor, the critical value $m$ is larger than at $\theta = \frac{2\pi}{n}$ (although still lower than at $\theta=0$). Thus, it is possible to interpret the generalized Haldane conjecture of [\onlinecite{2019arXiv191008196W}] in terms of these fractional topological excitations: When $p$ and $n$ are coprime, mass-generation only starts to occur for configurations that have winding number $\pm n$. In SU(2), these events are not strong enough to open a gap at the isotropic point $m=0$, and we predict that this holds for general $n$. In other words, we are claiming that the critical value of $m$ is zero when $p$ and $n$ are coprime. When $p$ and $n$ have a nontrivial common divisor different from $n$, configurations that have much less action begin to contribute to mass generation, starting with objects that have winding number $d$. The simplest example of this is in SU(4), with $p=2$. In this case, configurations that have winding number $\pm 2$ successfully generate a mass gap, while for $p=1$ these configurations destructively interfere, and a mass gap is not generated by the subleading configurations, with winding $\pm 4$. Finally, when $p$ is a multiple of $n$, the least action configurations that have winding $\pm 1$ produce a mass gap, just like the merons in SU(2).

While we haven't offered a rigorous argument as to why the critical value of $m$ is indeed fixed at zero in the case of $p$ and $n$ coprime, we can make the following observation: According to the LSMA theorem, a finite energy gap above the ground state implies spontaneously broken $\mathbb{Z}_n$ symmetry whenever $p$ is not a multiple of $n$. In the sG model  (\ref{eq:SG}), this $\mathbb{Z}_n$ symmetry corresponds to the following transformation:
\be \label{eq:zn2}
	\sigma \mapsto \sigma - \frac{2\pi}{n}.
\ee
While this transformation acts nontrivially on most fractional instantons, it has no effect on configurations with winding number $n$, which are the elementary excitations in the SG model when $p$ and $n$ are coprime. Thus, our prediction of a gapless phase in this case is perhaps not too unreasonable, since any finite gap would necessitate the spontaneous breaking of (\ref{eq:zn2}), which acts trivially on the Coulomb gas of $n$-winding events.

\section{Conclusions} \label{sec:con}

In this work, we have proposed a mass generating mechanism by fractional instantons in SU($n$) chains in the rank-$p$ symmetric representation. For $p$ a multiple of $n$, we have shown that topological configurations with charge $\frac{1}{n}$ produce a finite gap above the ground state, much in the same way that vortices produce a gap in the Kosterlitz-Thouless transition of the XY model. When $p$ and $n$ have a nontrivial common divisor different from $n$, we've shown that a mass gap is still produced, but now it is due to larger-action configurations that have charge $\frac{1}{\gcd(p,n)}$. Finally, when $p$ and $n$ are coprime, we have argued that no energy gap is produced by topological configurations. Together, these three results offer an intuitive explanation of the recent generalization of the Haldane conjecture to these SU($n$) chains.

\section*{Acknowledgements}

We would like to thank Pedro Lopes for helpful discussion. This work was supported by NSERC through Discovery Grant 04033-2016 and an NSERC PGS-D Scholarship, as well as by the QuEST Program of SBQMI.

\bibliography{meron.bib}
\bibliographystyle{apsrev4-1}

\appendix

\section{Explicit Form of Anisotropic Potentials on the SU($n$) Chain} \label{app:pot}

In this appendix, we determine what terms should be added to the SU($n$) chain Hamiltonian in order to give rise to the anisotropic potentials $V_1$ and $V_2$, defined in (\ref{eq:v}) and (\ref{eq:4b})  respectively. Using (\ref{eq:ss}), we have
\be
	|\phi_\alpha^\beta|^2 = \frac{1}{p} [S(j_\alpha)]_{\beta\beta},
\ee
so that
\be
	V_1 = \frac{m}{p^2} \sum_{\alpha=1}^n \sum_{\beta<\gamma} \Big( S(j_\alpha)_{\beta\beta} - S(j_\alpha)_{\gamma\gamma}\Big)^2.
\ee
and
\be
	V_2 = -\frac{m}{p^2} \sum_{\alpha=1}^n \sum_{\beta=2}^{n-1} \Big( [S(j_\alpha)_{1\beta}]^n - [S(j_\alpha)_{\beta 1}]^n\Big)^2.
\ee
To obtain the corresponding lattice terms, we simply multiply the RHS of these expressions by $n$, and sum over $j$. It is interesting to note that in the case of SU(3), the potential $V_1$ is proportional to
\be
	\sum_{\alpha=1}^3\left[T_3(j_\alpha) T_3(j_\alpha) + T_8(j_\alpha)T_8(j_\alpha)\right],
\ee
where $T_3$ and $T_8$ are the two diagonal generators of SU(3) in the Gell-Mann basis.

\section{Gauge Invariance of the Topological Charge} \label{app:1}

In (\ref{eq:12}), we rewrite the topological charge $Q$ as a difference of two contour integrals. It is interesting to note that while their sum is necessarily invariant under the $[\tU(1)]^{n-1}$ gauge symmetries (\ref{eq:3}), an individual contour integral is not. Indeed, under $\vec{\phi} \mapsto e^{i\theta}\vec{\phi}$, 
\be
	\oint dx^\mu \vec{\phi}\cdot\partial_\mu \vec{\phi}^*
	\mapsto 
	\oint dx^\mu \vec{\phi}\cdot\partial_\mu \vec{\phi}^*
	+ i \oint dx^\mu \partial_\mu \theta.
\ee
The second term, which breaks gauge invariance of the contour, is necessary in order to account for gauge transformations that alter the U(1) winding number of $\vec{\phi}$ along the contour. This feature implies that it is not meaningful to ask where the topological charge of a configuration $\vec{\phi}$ is localized. For instance, two gauge-equivalent versions of the SU(2) meron have the following behaviors at infinity:
\be
	\vec{\phi} = \frac{1}{\sqrt{2}}\begin{pmatrix} 1 \\
	e^{i\omega} \\
	\end{pmatrix} \hspace{10mm}
	\vec{\phi}' = \frac{1}{\sqrt{2}}\begin{pmatrix} e^{-i\omega} \\ 1 \\
	\end{pmatrix}.
\ee
Their behavior at zero is given by the first possibility in (\ref{eq:8}). The contour around infinity in (\ref{eq:12}) equals $-\frac{1}{2}$ for the case of $\vec{\phi}$, while it equals $+\frac{1}{2}$ for the case of $\vec{\phi}'$. Meanwhile, the contour around zero is trivial in the case of $\vec{\phi}$, and equals $+1$ for the case of $\vec{\phi}'$. The full expression in (\ref{eq:12}) yields $Q=\frac{1}{2}$ in both cases, as it must. Finally, we note that by choosing the gauge $\vec{\phi}' = (e^{-i\omega/2}, e^{i\omega/2})^\text{T}/\sqrt{2}$, the contour around infinity vanishes, which is what one finds in the O(3) version of the meron \cite{PhysRevLett.56.408}.

\section{An Explicit Fractional Instanton in SU(3)} \label{app:2}

Here, we prove the existence of a fractional instanton in SU(3), with topological charge $\vec{Q}  = (-\frac{2}{3}, \frac{1}{3},\frac{1}{3})$. While the most general complex unit vector in $\mathbb{C}^3$ depends on five real parameters, we make a symmetric ansatz that reduces this number down to two -- the minimum number of parameters necessary to construct a fractional instanton. For our three orthogonal vectors $\vec{\phi}^\alpha$, we take:
\be \label{eq:set1}
	\vec{\phi}^1 = \frac{1}{\sqrt{2}}\begin{pmatrix} \sin\alpha \\
	\sin\alpha \\
	\sqrt{2}\cos\alpha e^{i\omega} \\
	\end{pmatrix}
	\hspace{10mm}
	\vec{\phi}^2 = \frac{1}{2}\begin{pmatrix} \cos\alpha + i \\
	\cos\alpha -i  \\
	-\sqrt{2}\sin\alpha e^{i\omega} \\
	\end{pmatrix}
	\hspace{10mm} 
	\vec{\phi}^3 = \frac{1}{2} \begin{pmatrix} \cos\alpha - i \\
	\cos\alpha +i  \\
	-\sqrt{2}\sin\alpha e^{i\omega} \\
	\end{pmatrix}.
\ee
Here $\omega$ is the polar angle coordinate, and $\alpha\in [0,\pi/2]$ depends only on the radial coordinate: $\alpha = \alpha(r)$. Inserting this ansatz into the flag manifold sigma model Lagrangian (\ref{eq:2}), one obtains 
\be
	\mathcal{L} =  \frac{1}{g}\left[(\partial_r\alpha)^2 +\frac{1}{r^2}U_1(\alpha) + mg U_2(\alpha)\right],
\ee
where
\be
	U_1 (\alpha) = \sin^2\alpha\cos^2\alpha + \frac{1}{4}\sin^4\alpha,
\ee
and
\be
	U_2(\alpha) = \frac{5}{2}\left[ \cos^2\alpha - \frac{1}{2}\sin^2\alpha\right]^2 + \frac{1}{128}\cos^2\alpha(3\sin^4\alpha - 4\cos^2\alpha)^2.
\ee
Note that there is a unique coupling constant $g=g_{12}=g_{13}=g_{23}$ in SU(3), and the terms proportional to $\lambda_{\alpha\beta}$ have vanished. This is the classical Lagrangian for a particle at position $\alpha$ at time $r$, in an time-dependent effective potential
\be
	V_{\text{eff}}  = -\frac{1}{r^2}U_1(\alpha) - mgU_2(\alpha).
\ee

Now, at radii $r$ such that $r^2mg \ll 1$,  $V_{\text{eff}}$ is maximized by minimizing $U_1(\alpha)$ on $[0,\pi/2]$, which is achieved with $\alpha=0$. On the other hand, for $r\to \infty$, the effective potential is maximized by minimizing $U_2(\alpha)$, which is done with $\alpha = \alpha_0 = \arccos(1/\sqrt{3})$. Since $\frac{d}{d\alpha}U_1 \geq 0$ on $[0,\alpha_0]$, and $\frac{d}{d\alpha}U_2 \leq 0$ on the same interval, we may conclude that a solution exists with $\alpha(r)$ increasing monotonically from 0 at $r=0$ to $\alpha_0$ at $r=\infty$. As expected, the vectors $\vec{\phi}^\alpha$ in (\ref{eq:set1}) tend to (\ref{eq:typical}), up to a gauge transformation, as $r\to \infty$. Moreover, as $r\to 0$, the UV singularities $e^{i\omega}$ in the $\vec{\phi}^2$ and $\vec{\phi}^3$ vanish, and $\vec{\phi}^1 \to (0,0, e^{i\omega})^\text{T}$.

\end{document}